\documentstyle[11pt,aaspp4,flushrt]{article}

\slugcomment{Accepted for publication in ApJ}

\lefthead{M.~Kramer}
\righthead{Determination of the geometry of PSR 
B1913+16 by geodetic precession}

\begin{document}

\title{Determination of
the geometry of the PSR B1913+16 system by geodetic precession}

\author{Michael Kramer\altaffilmark{1}}

\affil{Astronomy Department, 601 Campbell Hall,
University of California, Berkeley, California 94720, USA,
email: michael@astro.berkeley.edu}
\altaffiltext{1}{also: Max-Planck-Institut 
f\"ur Radioastronomie, Auf dem H\"ugel 69,
53121  Bonn, Germany} 

\begin{abstract}
New observations of the binary pulsar B1913+16 are presented. Since 1978
the leading component of the pulse profile has weakend dramatically by
about 40\%. For the first time, a decrease in component separation is
observed, consistent with expectations of geodetic precession. Assuming
the correctness of general relativity and a circular hollow-cone like
beam, a fully consistent model for the system geometry is
developed. The misalignment angle between pulsar spin and orbital
momentum is determined giving direct evidence for an asymmetric
kick during the second supernova explosion. It is argued that the orbital
inclination angle is $132\fdg8$ (rather than $47\fdg2$). A prediction
of this model is that PSR B1913+16 will not be observable anymore
after the year 2025.
\end{abstract}
\keywords{pulsars: individual (B1913+16) - stars: neutron - supernovae:
- general}

\section{Introduction}

Following the discovery of the first binary pulsar B1913+16 (Hulse \&
Taylor 1975) it was shown that its unseen companion is another neutron
star (Taylor, Fowler \& McCulloch 1979). Although four other neutron
star binary systems are now known (see e.g.~Fryer \& Kalogera 1997 for
a recent discussion), PSR B1913+16 remains a
unique system in the sense that it exhibits a measurable amount of
geodetic precession (Damour \& Ruffini 1974; Barker \& O'Connell
1975a,b; Esposito \& Harrison 1975).
The effect occurs if the pulsar spin axis is misaligned with respect to
the orbital angular momentum vector. In this case, a coupling of the
vectors causes a precession of the pulsar spin around the orbital
momentum. As a result, the angle between spin axis and our line-of-sight
will change with time, so that different portions of the emission beam
are observed.  Consequently, one expects changes in the measured pulse
shape as a function of time.  Weisberg, Romani \& Taylor (1989,
hereafter WRT89), reported a change in the relative amplitude of the
prominent leading and trailing components in the pulse profile of PSR
B1913+16. Analyzing observations made at 1408 MHz in a time interval
from 1981 to 1987 they determined a change in the amplitude ratio of
$1.2\pm0.2$\% yr$^{-1}$, leading to a secular weakening of the leading
component.  If the emission beam exhibits an overall hollow-cone shape,
one would also expect a change in the {\em separation} of the two
components rather than only a change in relative intensity. This was,
however, not observed within an upper limit of $0\fdg06$ for the
given time interval. As an interpretation, WRT89 argue that geodetic
precession alters the observer's cut through an irregular and patchy
emission beam structure, similar to that proposed by Lyne \& Manchester
(1988).  

Cordes, Wasserman \& Blaskiewicz (1990, hereafter CWB90) used
polarization data of PSR B1913+16 to compare waveforms and position angle
(PA) swings which were obtained at slightly different frequencies
between 1397 and 1416 MHz between 1985 and 1988.  In the classical
polar cap model the PA swing depends on the impact parameter of our
line of sight (Radhakrishnan \& Cooke 1969), so that geodetic
precession should alter the swing.  While CWB90 detected a only
possible change in the waveform, they could not find any significant
change in the PA swing.  In order to reconcile their observations with
those of WRT89, they argue that while geodetic precession is acting,
the existence of a core component, which is more prominent at lower
frequencies (cf.~Taylor et al.~1979), is affecting the otherwise
uniform hollow cone structure and the stability of the component
ratio.

In the following we present new observations of PSR B1913+16 and propose a
model giving an estimate of the misalignment angle between pulsar
spin axis and orbital momentum.  Such angle is particularly useful
to constrain the progenitor system and the asymmetry of the second
supernova explosion (e.g.~Cordes \& Wasserman 1984, Burrows \& Woosley
1986).
In Sec.~\ref{obs} we describe our observations and the analysis
of the data which we model in Sec.~\ref{model}. We discuss our
results in Sec.~\ref{discus} and draw our conclusions in 
Sec.~\ref{conclusions}.

\section{Observations and data analysis}

\label{obs}

The observations were made on six epochs between April 1994 and April
1998, using the 100-m Effelsberg radiotelescope of the
Max-Planck-Institut f\"ur Radioastronomie. We used a tunable
1.3/1.7-GHz-HEMT receiver installed in the prime focus yielding a system
temperature of 25--35 K on the cold sky with a telescope gain of 1.5
K/Jy. The system provided left and right hand circularly polarized
signals which were centered on 1410 MHz.  Full detail of the observing
system can be found in Kramer et al.~(1998).

Apart from a variation in observing time between 30 to 60 min depending
on interstellar scintillation, each of the incoherently de-dispersed
data sets was obtained and analyzed in a homogeneous way.  Total power
profiles were produced by gain calibrating and summing the two
circularly polarized intensities. These profiles were smoothed to match
the dispersion smearing across one filterbank channel, resulting in an
effective time resolution of 337 $\mu$s compared to a 
sampling time of 57.6 $\mu$s (cf.~Fig.~\ref{profile}). The effective
time resolution of our data is about a factor of 2.5 worse than that of
the data presented by WRT89, but about a factor of 1.6 better than
CWB90. Since our data are at the same center frequency (i.e.~within 0.1\%)
and have similar time resolution as WRT89 and CWB90, we can
accurately compare our data set to the previous observations. 

At first, we measured the amplitude ratio of the leading and trailing
components using the off-pulse RMS as an uncertainty estimator.  These
are presented in Fig.~\ref{ratio}, where we also plot data from WRT89.
It is seen that the new data are in good agreement with the earlier
measured trend. Performing a weighted least-squares fit, we obtained a
ratio change of $-1.9\pm0.1$\% per year.  It is interesting to note
that our fit is also in agreement with the component ratio from March
1978 (Fig.~\ref{ratio}), which was measured using the profile published
by Taylor \& Weisberg (1982).  The amplitude ratio between profile
midpoint and trailing component seems to remain constant at a value of
about 0.17 over a time span of 20 years within the errors.

Measuring the separation between the leading and trailing components,
we followed the approach of WRT89 by fitting Gaussian curves to the
central portion of each component. The formal accuracy determined by
the fit error can be up to an order of magnitude better than the
sampling time, similar to a standard template matching procedure. We
note, however, that the components are asymmetric in shape, so that
the resulting centroids might be affected by the number of samples
included in the fit around the peak.  In order to estimate this effect
as the true separation uncertainty, we performed fits including
samples corresponding to intensity levels of down to 95\%, 90\%, 80\%
and 60\%, respectively.  The resulting mean values for the separation
and their corresponding errors are presented in Fig.~\ref{widths} with
values adapted from WRT89. In order to verify this approach we
measured the component separation of the 1981 profile presented by
WRT89 at corresponding intensity levels. We obtained a value of
$38\fdg6\pm0\fdg1$, in very good agreement with their value.
This gives confidence in the consistency between our measurements
and those of WRT89.

\section{Modeling geodetic precession}

\label{model}

In what follows we will make two assumptions: a) general relativity is
the correct theory of gravitation within the uncertainties of our
measurements; and b) the emission beam exhibits an overall circular
hollow-cone like shape, with intensity possibly depending on magnetic
longitude.  The first assumption is obviously well justified, given the
excellent agreement of the measured orbital decay with the prediction
from general relativity (Taylor \& Weisberg 1982, 1989). While the assumption
of an intensity dependence on magnetic longitude accounts for the
component ratio change, the hollow-cone like shape follows the arguments
given by CWB90, which can now be based on even more data presented for
slowly rotating pulsars (e.g.~Rankin 1993, Gil et al.~1993, Kramer et
al.~1994, Gould 1994).  These data, derived from obviously random cuts
through emission beams, indicate that the opening angle of the beam,
$\rho$, follows a period relation as expected for a circular beam in a
dipolar field structure.  Recent results suggest that this is not true
anymore for very fast rotating pulsars ($P\lesssim40$ms) but still
applicable to PSR B1913+16 (Kramer et al.~1998). 

Assumption (a) and the timing parameters presented by Taylor (1992) imply
an orbit inclination angle, $i=47\fdg2$, or
$i=180\arcdeg-47\fdg2=132\fdg8$, and an expected precession rate
of $\Omega_{\rm p}=1.21\arcdeg$ yr$^{-1}$ (Barker \& O'Connell 1975a,b).
Modeling the effects of precession we use the same notation and
coordinate system as introduced by CWB90 but define the precession phase
\begin{equation} 
\phi(t)=\Omega_{\rm p}\cdot(T_0-t) 
\end{equation} 
such that $T_0$ describes the closest approach of the pulsar spin axis,
$\bf \Omega$, to the line-of-sight, $\bf n$ (see Fig.~\ref{coordsys}).
Note that our coordinate system represents the usual observers practice
(i.e.~pulsar-centered with z-axis anti-parallel to the angular momentum
vector) and thus differs from that described by Damour \& Taylor (1992),
which is Earth-centered with the primary axis along the line-of-sight.
For a comparison, in Fig.~\ref{coordsys} we refer to Damour \& Taylor's
definition of $\bf K$, $\bf k$ and $\bf s_{\rm 1}$ (cf.~their Fig.~1).
Despite this difference, the misalignment angle, $\lambda$, between
pulsar spin and orbital momentum is the same in both their and our
convention. The angle between pulsar spin and line-of-sight, $\beta$,
will change with time according to
\begin{equation} 
\cos \beta(t) = \cos \lambda \, \cos i + \sin \lambda \, \sin i \, \cos \phi(t).  
\end{equation} 
affecting also
the impact parameter, $\sigma(t)=\beta(t)-\alpha$, i.e.~the angle
between the magnetic axis, $\bf m$, and the line-of-sight at their
closest approach.  If the inclination angle between magnetic axis and
pulsar spin axis is denoted by $\alpha$ and the emission beam exhibits
an opening angle, $\rho$, the observed pulse width (or component
separation if $\rho$ is defined accordingly) changes with time as 
\begin{equation} 
W(t) = 4\sin^{-1} \left[ \sqrt{\frac{\sin^2 (\rho/2) -
\sin^2(\sigma(t)/2)}{
       \sin\alpha \sin \beta(t)}} \right]
\end{equation}
using assumption (b).  Obviously, only four free parameters describe
the system discussed here, i.e.~$\alpha$, $\lambda$, $T_0$ and
$\rho$. In order to determine them, we fitted the measured component
separations to the equations above. First we started a robust Simplex
least-squares fit algorithm (e.g.~Caceci \& Cacheris 1984) on grid
positions covering the whole parameter space (i.e.~$0\le\alpha$,
$\lambda\le180$, $1997-150\le T_0\le 1997+150$, $0\le\rho\le 90$),
treating $\rho$ as a free parameter. However, due to WRT89's very
precise measurements, $\rho$ can be essentially fixed after
calculating $\beta$ and $\sigma$ for chosen values of $\alpha$,
$\lambda$, $T_0$ and a component separation measured in 1981. The
$\chi^2$-hypersphere was thus searched on a finer grid in the
remaining three dimensions. Both methods yield the same results
with a best reduced $\chi^2_{\rm red}=6.94$ and 8
degrees-of-freedom. This relatively large $\chi^2$-value
suggests that some of the separation uncertainties are underestimated.
Nevertheless, the minimum is well constrained as shown by the 
68\%, 90\% and 99\% confidence contours (Fig.~\ref{contours})
and confirmed by Monte-Carlo simulations.  However, since
the orbit inclination angle can be either $i=47\fdg2$ or
$i=132\fdg8$, four equivalent solutions exist which are
listed in Table~\ref{solutions} (note that the angles 
are related to each other by a
180$\arcdeg$-ambiguity). Quoted uncertainties represent 99\% confidence
limits. As we discuss in detail below, we consider solution III as
the correct one.

\section{Discussion}
\label{discus}

We observe a gradual decrease in the amplitude of the leading component, 
which is apparently caused by geodetic precession
(cf.~WRT89, CWB90).  While the amplitude ratio of the trailing component
to the profile midpoint seems to remain unchanged, flux density
variations due to interstellar scintillation prevent a conclusion
whether the profile has weakened as a whole. Nevertheless, the observed
amplitude change strongly suggests a latitudinal dependence of the
emission strength.  This is not too surprising as, e.g., dependencies of
the spectral index on the distance to the magnetic axis have been
observed before (e.g.~Lyne \& Manchester 1988, Kramer et al.~1994). The
case of PSR B1913+16 is however unique as such dependence can be seen for a
single pulsar.  Assuming the model parameters in Table~\ref{solutions},
the impact parameter has changed by about $2\arcdeg$ between 1981 and
1998. This corresponds to about 20\% of the determined opening angle,
which seems a realistic value for the given amplitude change.
Although the beam is thus not uniformly filled, this does not contradict
the assumption about a circular beam shape.  In fact, CWB90 show
that the core component which lags the profile midpoint significantly,
can disturb a possible uniformity of the beam.  Taylor et al.~(1979) had
argued similarly based on reported profile changes at 430 MHz. Later,
however, they attributed the effect to a comparison of data obtained
with a different number of polarization channels (Taylor \& Weisberg
1982).  We thus believe that the simple assumption of a circular hollow-cone
like beam shape is well justified. 

Since we have determined the system geometry without using polarization
data, we can actually compare our results with those determined using
polarization observations by CWB90 and Blaskiewicz, Cordes, \&
Wassmerman (1991, hereafter BCW91). Using the same observer's
convention as CWB90 and BCW91, we can make this comparison directly
without a necessary transformation of the involved angles as it would be
necessary when using Damour \& Taylor's (1992) 
convention. Computing the impact parameter $\sigma$ for the observing
epochs of CWB90 and BCW91, we note that solutions I and IV yield a
negative value, while solutions II and III give a positive impact angle.
Both CWB90 and BCW91 present PA curves which are well defined over a wide
range of longitudes ($\gtrsim 60^\circ$), exhibiting a positive slope
with a flattening at pulse longitudes separated from the profile midpoint 
by more than $15^\circ$. This behaviour clearly indicates a positive
impact parameter, i.e. an outer line-of-sight (Narayan \& Vivekanand 1982,
Lyne \& Manchester 1988), so that we can
exclude the first and fourth solutions. The remaining ones differ in
particular in the value of the misalignment angle.  While solution III
indicates an almost aligned system, solution IV implies almost
anti-parallel spins. Such a retrograde spin requires an asymmetric kick
during the second supernova explosion which has to be much greater 
than the pre-supernova orbital velocity of the pulsar
(e.g.~Hills 1983, Kaspi et al.~1996).  
Following the
calculations of Hills (1983) or Freyer \& Kalogera (1997), this would
imply a kick velocity of about 500 km s$^{-1}$, which is in conflict with
Freyer \& Kalogera's estimate of about 200 km s$^{-1}$ as the most probable
value. Such large kicks are also hard to reconcile with the (transverse)
system velocity of PSR B1913+16 of about 130 km s$^{-1}$
(Taylor \& Weisberg 1989).  Hence, we reject this solution
and consider model III as the one describing the data best.

Although we are confident in the consistency between our and WRT89's data,
we now discuss the possibility that the comparison of data which were
obtained with different observing systems, can bear some
uncertainties. Obviously, we minimized possible effects by using very
similar observing parameters and an identical data analysis procedure
as WRT89. We can nevertheless study remaining uncertainties assuming
even larger measurement errors than estimated. The best solutions
listed in Table~\ref{solutions} predict that the PA swing steepens
with time, while a sign change in the impact parameter will occur in
the near future (or has already occured recently). With increased
measurement errors, solutions are allowed
(e.g.~$\lambda\approx22\arcdeg$, $\alpha\approx112\arcdeg$) where the
impact parameter has changed sign prior to the observations of CWB90,
so that it will remain positive with less steep PA swings.  Studies of
various $\chi^2$-hyperspheres, however, indicate that the most
interesting parameter, $\lambda$, will most likely remain in a range
$13\arcdeg \lesssim\lambda \lesssim25\arcdeg$.

Using model III to compute the change in the impact angle, $\sigma$,
between CWB90's observing epochs, we derive a change of
$\Delta\sigma=\sigma(1988.5)-\sigma(1984.5)=0\fdg8 - 1\fdg1 =
-0\fdg3$, which is only slightly higher than their estimated detection
limit.  Following also the discussion by CWB90 that the core component
affects the PA curve, it is thus not surprising that they were not
able to detect a change.  In fact, the authors argue that the impact
parameter should be much smaller than implied by their measured slope
of the PA swing, which is also consistent with our result.  Following
the arguments of CWB90, BCW91 excluded the central portion of the PA
curve, when they fitted a (relativistic) Radhakrishnan-Cooke (1969)
model to measured PAs of PSR B1913+16. At 1403 MHz they obtained
$\alpha=171\arcdeg\pm64\arcdeg$ and $\sigma=0\fdg2\pm1\arcdeg$, which
is indeed in good agreement with model III.

Our results imply a misalignment of the pulsar spin to the orbital
momentum vector of $\lambda=22^{+3}_{-8}\arcdeg$. 
Such value is in agreement with early
simulations made by Bailes (1988) who found $20\arcdeg$ as a typical
value for PSR B1913+16 like systems. The model also implies that the
component separation remains almost unchanged for about 60 yr, which
corresponds to a likelihood of 20\% to observe 
the pulsar in that phase given a
precession period of about 300 yr. Thus, it is easy to understand why
WRT89 did not detect any change in the component separation.

\section{Conclusions}

\label{conclusions}

We have analyzed new observations of PSR B1913+16, showing that the
first component has weakened by about 40\% since 1978. For the first
time we detected a small change in component separation, which we modeled
assuming geodetic precession and a circular hollow-cone like beam.
If our model is correct, the change of the component separation should
be easily measurable in future high signal-to-noise ratio observations
as it will decrease with a higher rate. A test is also provided by
monitoring the PA curve whose slope will possibly change sign in the near
future. 

Due to geodetic precession, the pulsar will not beam towards Earth for
a significant fraction of its lifetime.  In fact, our proposed model
predicts that PSR B1913+16 will not be observable anymore after the
year $\sim$2025. Shortly before this, the leading component will
disappear if it continues to weaken with the measured rate.  
Reappearing again around the year 2220, PSR B1913+16 will, in total,
only be observable for about a third of the precession period
(cf.~Fig.~\ref{widths}). While this does not affect birth rate
calculations of double neutron star systems, the relatively small
inferred (two-pole) beaming fraction of about 0.35 should be
considered in future calculations. 

So far, we have assumed that general relativity is the correct theory
of gravitation in order to calculate the precession rate.
With more and better data, we can  treat the precession rate
as a free parameter, so that the performance of another test for general 
relativity will be possible.

While timing observations are not able to decide between the ``true''
orbital inclination angle, our final model suggests that its value is
$i=132\fdg8$ rather than $i=47\fdg2$. While polarization data
provide information of the system in the two dimensions of the
projected sky, geodetic precession and its observed effects yield
information in direction of our line-of-sight. Combining all
the available information allows to draw a three dimensional
picture and to the remove the ambiguity in the orbital inclination
angle of PSR B1913+16 for the first time.

It is clear that a misalignment angle of $\lambda=22^{+3}_{-8}\arcdeg$ 
requires an asymmetric kick during the second
supernova explosion. An asymmetric kick has been already inferred for
J0045$-$7319 (Kaspi et al.~1996) and has been argued for B1913+16 on
basis of evolutionary models (e.g.~Wijers et al.~1992, van den
Heuvel \& van Paradijs 1997, Fryer \& Kalogera 1997). Our observations
provide direct evidence for this hypothesis and give a natural
explanation for the large inferred birth verlocities of pulsars
(e.g.~Lyne \& Lorimer 1993).

\acknowledgments

I am indebted to all people involved in the project to monitor millisecond
pulsars in Effelsberg, in particular to Kiriaki Xilouris, Christoph
Lange, Dunc Lorimer and Axel Jessner.  During this work I enjoyed
stimulating discussions with Norbert Wex and Duncan Lorimer, who also read
the manuscript with great care. It is a particular pleasure to thank Don 
Backer for the countless fruitful and helpful discussions. 
I am grateful for the constant support of Richard Wielebinski, and
acknowledge the receipt of the Otto-Hahn Prize, during
whose tenure this paper was written.

\clearpage

\figcaption{Pulse profile of B1913+16 observed in April 1995 \label{profile}}

\figcaption{Amplitude ratio of leading and trailing components as
a function of time. Filled circles were adapted from WRT89, while
open circles represent Effelsberg measurements. The dashed line
represents a linear fit to the data with a slope of 1.9\% yr$^{-1}$.
The amplitude ratio adapted from TW82 is shown as a filled
square
\label{ratio}}

\figcaption{Measured component separations as a function of time.
Closed circles are values presented by WRT89, while open circles
are measurements presented here. The solid line represents the
best fit models listed in Table 1. The open square shown in the upper
panel demonstrates the consistency of our measurement method with that
of WRT89 (see text for details) \label{widths}}

\figcaption{Coordinate system to describe geodetic precession (see text
for details). 
Vectors $\bf K$, $\bf k$ and $\bf s_{\rm 1}$
refer to Fig.~1 of Damour \& Taylor (1992) \label{coordsys}}

\figcaption{Contour plots of 66\% (solid line), 90\% (dashed line) and 99\% 
(dotted line) confidence limits
around the solution of model III. \label{contours}}

\clearpage
 
\begin{deluxetable}{rccccc}
\footnotesize
\tablecaption{Best model solutions discussed in the text
\label{solutions}}
\tablewidth{0pt}
\tablehead{
 & \colhead{$i$} & \colhead{$\alpha$}   & \colhead{$\lambda$} 
& \colhead{$T_0$} & \colhead{$\rho$} \\
  &  \colhead{(deg)} & \colhead{(deg)} & \colhead{(deg)} & \colhead{(yr)}
  & \colhead{(deg)} }
\startdata
I. &  \phn 47.2 & \phn$27^{+8}_{-3}$ & \phn$22^{+3}_{-8}$ & $1980\pm4$ & 
$9.0^{+4}_{-3}$ \nl
II. & \phn 47.2 & $153^{+3}_{-8}$ & $158^{+8}_{-3}$ & $2128\pm4$ & 
$9.0^{+4}_{-3}$ \nl
III. & 132.8 & $153^{+3}_{-8}$ & \phn$22^{+3}_{-8}$ & $2128\pm4$ & 
$9.0^{+4}_{-3}$ \nl
IV. & 132.8 & \phn$27^{+8}_{-3}$ & $158^{+8}_{-3}$ & $1980\pm4$ & 
$9.0^{+4}_{-3}$ \nl
\enddata
\end{deluxetable}

\clearpage

\plotone{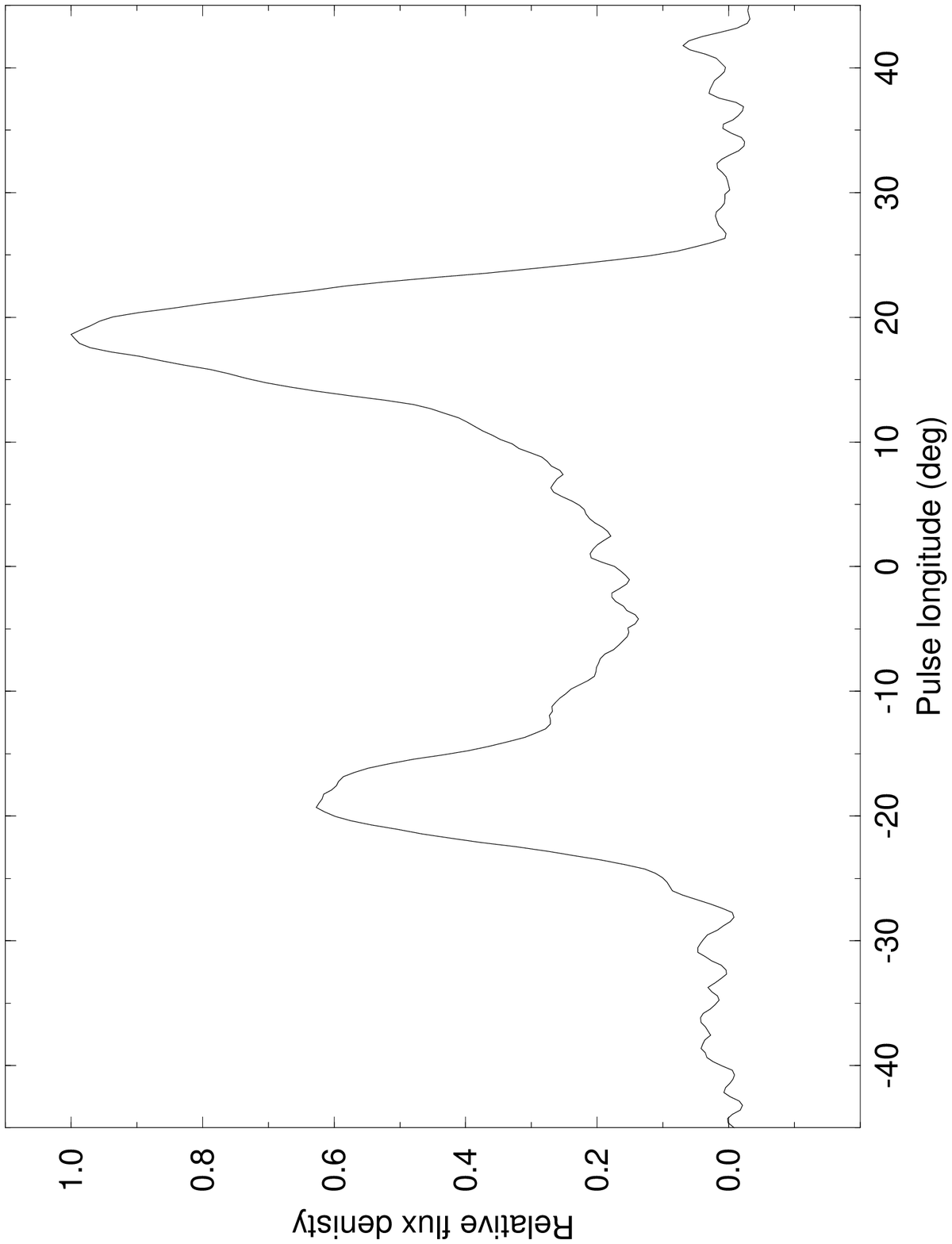}


\plotone{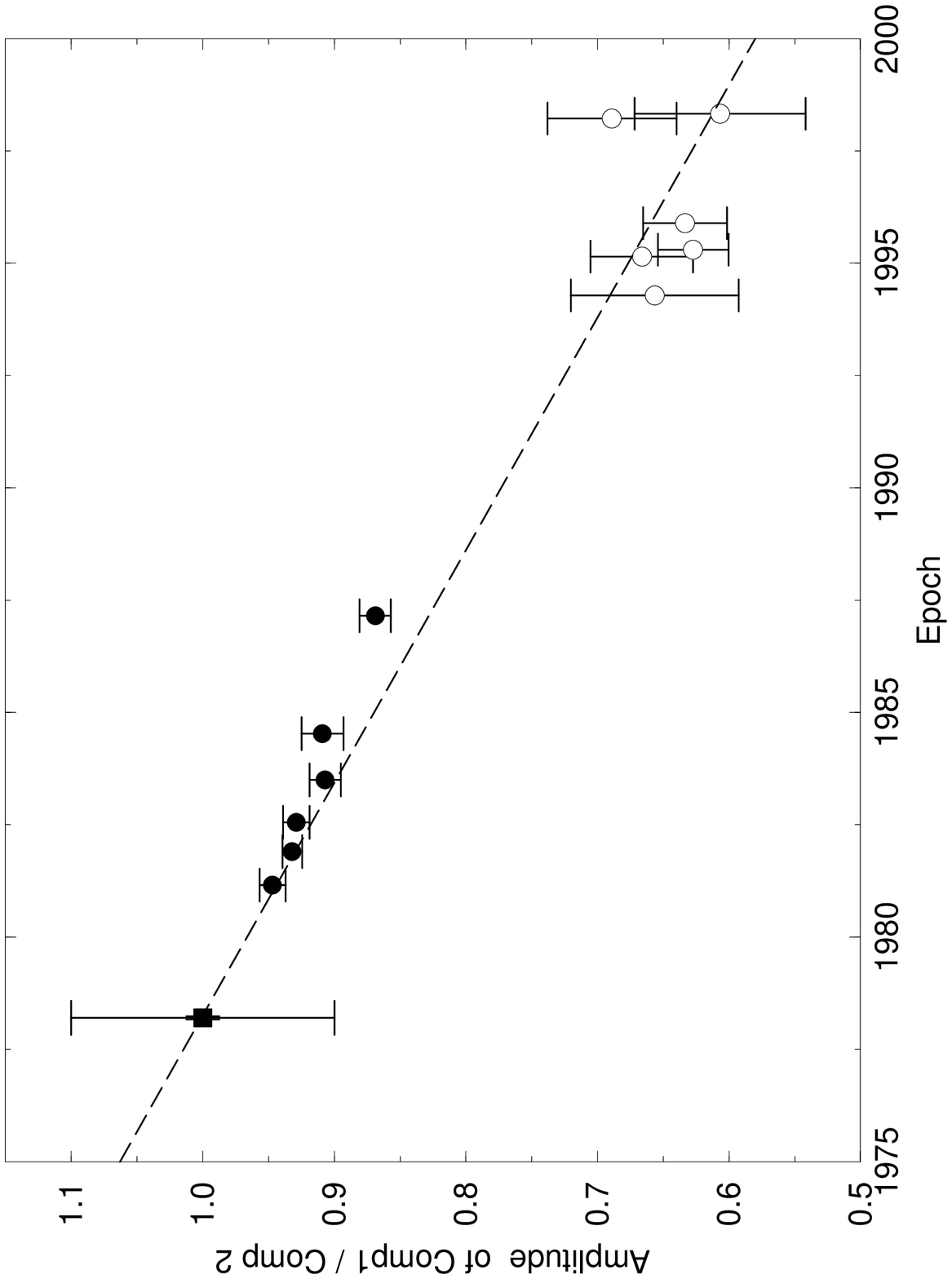}


\plotone{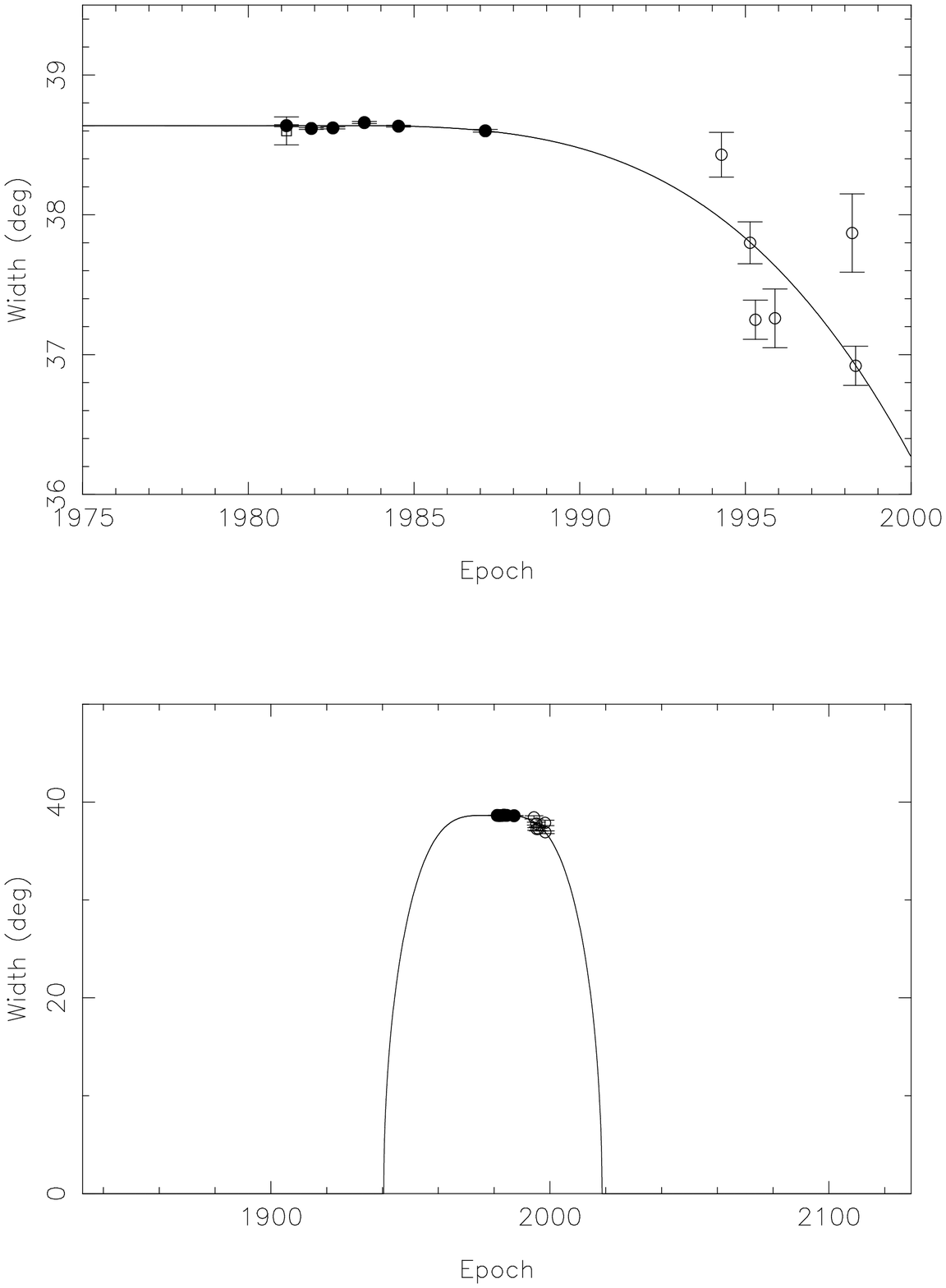}


\plotone{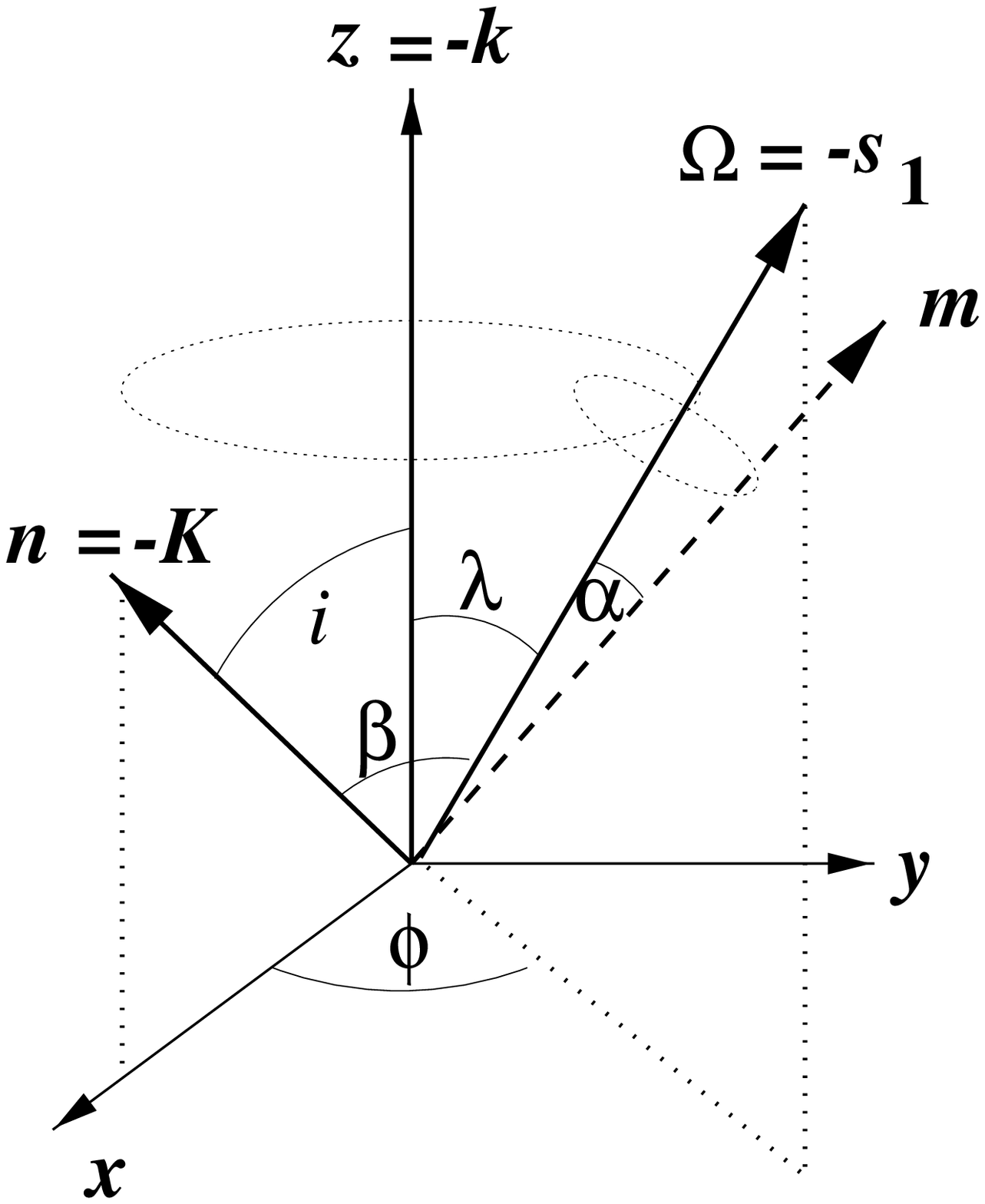}


\plotone{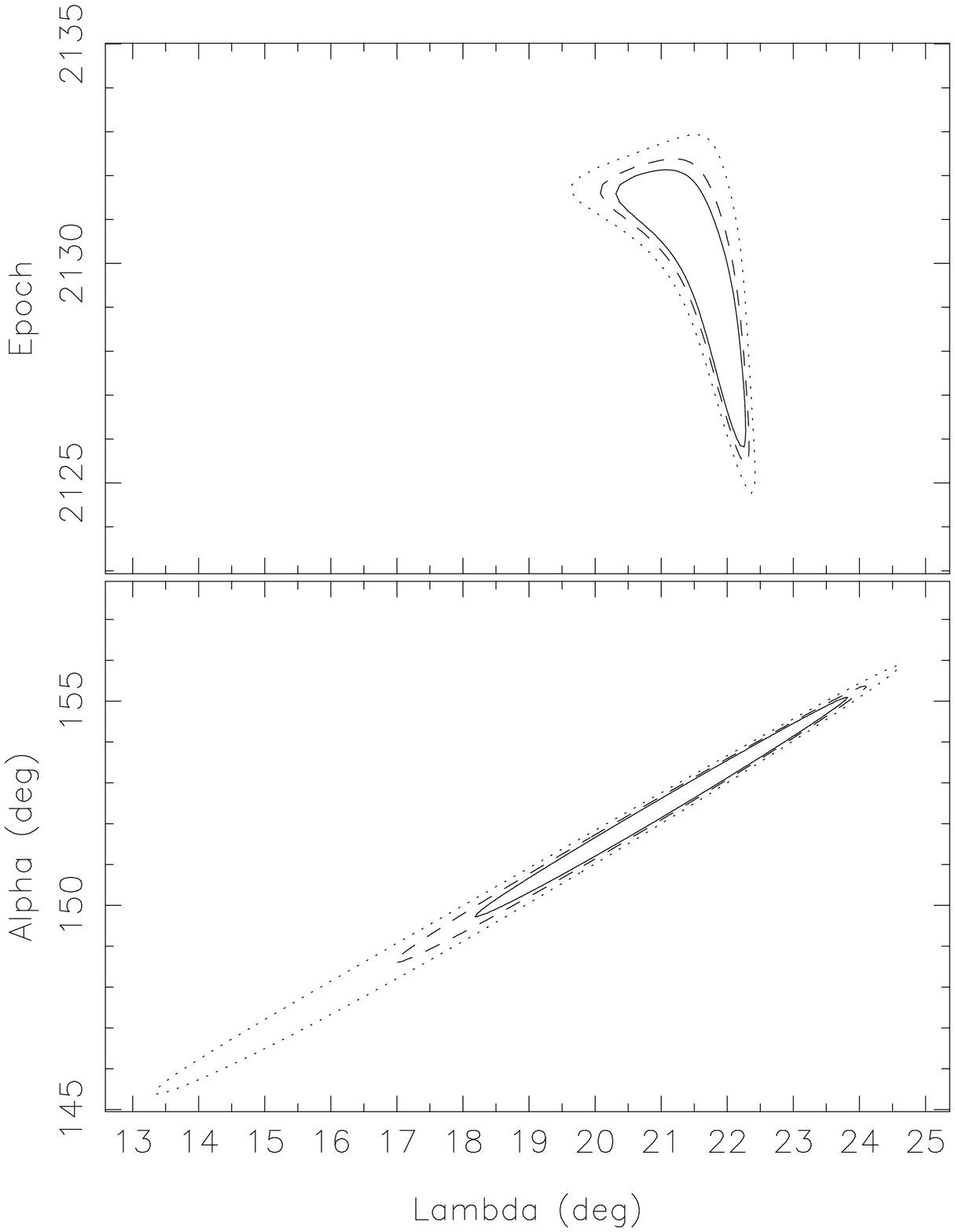}


\begin{thebibliography}{DUM}
%
\bibitem{bc75a} Barker, B.M., \& O'Connell, R.R.~1975a, \prd, 12, 329 
%
\bibitem{bc75b} Barker, B.M., \& O'Connell, R.R.~1975b, \apj, 199, L25
%
\bibitem{bal88} Bailes, M.~1988, \aap, 202, 109 
%
\bibitem{bcw91} Blaskiewicz, M., Cordes, J.M., \& Wasserman, I.~1991,
\apj, 349, 546 (BCW91)
%
%
\bibitem{bw89} Burrows, A., \& Woosley, S.~1986, \apj, 308, 680 
%
\bibitem{cc84} Caceci, M.S., \& Cacheris, W.P.~1984, Byte, 5, 340
%
\bibitem{cs84} Cordes, J.M., Wasserman, I., 1984, \apj, 279, 798 
%
\bibitem{cwb90} Cordes, J.M., Wasserman, I., \& Blaskiewicz, M.~1990, 
\apj, 349, 546 (CWB90)
%
%
\bibitem{dr75} Damour, T., \& Ruffini, R.~1974, \prd, Acad.~Sci.~Paris,
279, s{\'e}rie A, 971
%
\bibitem{dt92} Damour, T., \& Taylor, J.H.~1992, \prd, 45, 1840
%
\bibitem{eh75} Esposito, L.W., \& Harrison, E.R. 1975, \apj, 196, L1 
%
\bibitem{fk97} Fryer, C., \& Kalogera, V.~1997, \apj, 489, 244 
%
 \bibitem[Gil {\it et al.} 1993]{gks93} 
 Gil, J.~A., Kijak, J., \& Seiradakis, J.~H.~1993, \aap, 272, 268
%
\bibitem[Gould (1994)]{gou94}
Gould, D.~M.~1994, PhD-thesis, University of Manchester
%
\bibitem{H83}
Hills, J.G.~1983, \apj, 267, 322
%
%
\bibitem{ht75} Hulse, R.A., \& Taylor, J.H.~1975, \apj, 195, L51
%
\bibitem{kbm+96} Kaspi, V.M., Bailes, M., Manchester, R.N., Stappers,
B.W., Bell, J.F.~1996 \nat, 381, 584
%
  \bibitem[Kramer {\it et al.}  1994] 
  {kra94} Kramer, M., Wielebinski, R., Jessner, A., Gil, J.~A., \&
  Seiradakis, J.~H.~1994, \aaps, 107, 515
%
%
  \bibitem[Kramer {\it et al.}  1998] 
  {kra98} Kramer, M., Xilouris, K.M., Lorimer, D.R, Doroshenko, O.,
Jessner, A., Wielebinski, R., Wolszczan, A., \& 
Camilo, F., 1998, \apj, 501, in press
%
\bibitem{lm88} Lyne, A.G., \& Manchester, R.N.~1988, \mnras, 234, 477
%
\bibitem{ll93} Lyne, A.G., \& Lorimer, D.R.~1993, \nat, 369, 127
%
\bibitem{nv82} Narayan, R., \& Vivekanand, M.~1982, \aap, 113, L3
%
\bibitem{rc69} Radhakrishnan, V., \& Cooke, D. J.~1969, Astrophys.~Lett., 3, 225
%
 \bibitem[Rankin 1993]{R93} Rankin, J.~M.~1993, \apj, 405, 285
%
%
\bibitem{tay92}Taylor, J.H.~1992, Phil.~Trans.~R.~Soc.~Lond.~A, 341, 117
%
\bibitem{tfc79} Taylor, J.H., Fowler, L.A., \& McCulloch, P.M., 1979, 
\nat, 277, 437
%
%
\bibitem{tw82} Taylor, J.H., Weisberg, J.M.~1982, \apj, 253, 908
%
\bibitem{tw89} Taylor, J.H., Weisberg, J.M.~1989, \apj, 345, 434
%
\bibitem{wrt89} Weisberg, J.M., Romani, R., \& Taylor, J.H.~1989, \apj,
347, 1029 (WRT89)
%
\bibitem{wph92} Wijers, R.A.M.J., van Paradijs, J., \& van den Heuvel, E.P.J.~1992, \aap, 261, 145
%
\bibitem{hp97} van den Heuvel, E.P.J, \& van Paradijs, J.~1997, \apj, 483, 399
%
\end{thebibliography}
\end{document}